\documentclass[aps,twocolumn,pra,longbibliography]{revtex4-2}
\PassOptionsToPackage{hyphens}{url}

\usepackage[T1]{fontenc}
\usepackage{amsmath,amssymb,amsfonts}
\usepackage{graphicx}
\usepackage{xcolor}
\usepackage{textcomp}
\usepackage{placeins}

\usepackage[ruled,vlined]{algorithm2e}
\usepackage{subfig}

\begin{document}

\title{\textsc{MoSAIC}: Scalable Probabilistic Error Cancellation via Variational Blockwise Noise Aggregation}

\author{Maya Ma}
\affiliation{Department of Computer Science, University of California Santa Barbara}
\email{yema@ucsb.edu}

\author{Rimika Jaiswal}
\affiliation{Department of Physics, University of California, Santa Barbara}
\email{rimika@ucsb.edu}

\author{Murphy Yuezhen Niu}
\affiliation{Department of Computer Science, University of California, Santa Barbara}
\email{murphyniu@ucsb.edu}

\begin{abstract}
Quantum error mitigation is essential for extracting trustworthy results from noisy intermediate-scale quantum (NISQ) processors. Yet, current approaches face a core scalability bottleneck: unbiased methods such as probabilistic error cancellation (PEC) incur exponential sampling overhead, while approximate techniques like zero-noise extrapolation trade accuracy for efficiency. We introduce and experimentally demonstrate \textsc{MoSAIC} (Modular Spatio-temporal Aggregation for Inverted Channels), a scalable quantum error mitigation framework that preserves the unbiasedness of PEC while dramatically reducing sampling costs. \textsc{MoSAIC} partitions a circuit into noise-aligned blocks, learns an effective block noise model using classical variational optimization, and applies quasi-probabilistic inversion once per block instead of after every layer. This blockwise aggregation reduces both sampling overhead and circuit-depth overhead, enabling mitigation far beyond the operating regime of standard PEC. We also experimentally validate \textsc{MoSAIC} on IBM's 156-qubit Heron processors, performing the largest PEC-based mitigation demonstration on hardware to date. As a physically meaningful benchmark, we prepare the critical one-dimensional transverse-field Ising (TFIM) ground state for system sizes up to 50 qubits. We show that \textsc{MoSAIC} can achieve at least 1-2 orders of magnitude better accuracy than standard PEC under identical sampling budgets. This enables \textsc{MoSAIC} to recover accurate observables for larger system sizes, even when standard PEC fails due to its prohibitive sampling overhead. We also present CUDA-Q accelerated simulations to validate performance trends under a range of different noise models. These results demonstrate that \textsc{MoSAIC} is not only theoretically scalable but also practically deployable for high-accuracy, large-scale quantum experiments on today’s quantum hardware.
\end{abstract}

\maketitle

\section{Introduction}

Quantum computers are expected to offer computational advantages over classical machines in applications such as quantum chemistry, materials simulation, and optimization. Rapid progress in hardware development has steadily increased qubit counts and improved gate fidelities, yet present-day devices remain far from the regime required to implement large-scale fault-tolerant algorithms. Even at moderate circuit depths, accumulated noise can quickly overwhelm the signal, causing measured observables to drift from their ideal values. As a result, scalable error-mitigation methods are essential for extracting useful information from near-term quantum hardware~\cite{Endo_2018,Cai_2023}.

Among existing error-mitigation techniques, \emph{probabilistic error cancellation (PEC)} is one of the most general and theoretically rigorous approaches for obtaining unbiased estimates of quantum observables~\cite{van_den_Berg_2023,PhysRevLett.119.180509,PRXQuantum.3.040313,PhysRevA.109.062617,Piveteau_2022}. PEC works by learning a model of device noise and constructing a quasi-probability representation of the inverse noise channel using physically realizable operations. By sampling from this distribution and reweighting measurement outcomes, one can in principle recover unbiased estimators of ideal expectation values. The main obstacle is that PEC becomes prohibitively expensive as circuits grow. Its sampling cost increases exponentially with circuit depth, noise strength, and system size in order to maintain fixed estimator variance. This scaling makes standard PEC increasingly impractical precisely in the large-circuit regime where useful quantum advantage is expected to emerge.

\begin{algorithm*}[th!]

\LinesNotNumbered
\caption{\textsc{MoSAIC} blockwise probabilistic error cancellation}
\label{algo:mosaic_recipe}
\DontPrintSemicolon
\small

\textbf{Input:} transpiled circuit $C$, observable $\hat O$, block grain $w$, Monte Carlo sample count $N$, shots per executed circuit $S$ \\
\textbf{Output:} mitigated estimate $E \approx \langle \hat O \rangle$

\vspace{0.5em}
\textbf{1. Choose block size (user-tunable).}
Select a grain $w$ (or a heterogeneous schedule $\{w_b\}$) to trade classical processing cost against QPU sampling cost.

\vspace{0.4em}
\textbf{2. Partition the circuit.}
Partition $C$ into an ordered list of blocks $\mathcal{B}=\{b_1,\dots,b_M\}$ using noise scope and locality at grain $w$.

\vspace{0.4em}
\textbf{3. Characterize and invert each block (classical preprocessing).}
For every block $b \in \mathcal{B}$:
\begin{enumerate}
\item[(a)] Obtain layer-level noise models within $b$ (for example Pauli--Lindblad channels from noise learning).
\item[(b)] \textbf{MoSAIC novelty: variational block noise learning.}
Fit a single effective block noise channel $\mathcal{N}_b$ that best matches the composed noisy block dynamics.
\item[(c)] Compute a quasi-probabilistic representation of a block recovery channel
$\mathcal{R}_b \approx \mathcal{N}_b^{-1} = \sum_r \eta_{b,r}\,\widetilde{\mathcal{O}}_{b,r}$,
where $\widetilde{\mathcal{O}}_{b,r}$ are implementable operations (typically Pauli corrections).
\item[(d)] Convert $\{\eta_{b,r}\}$ into sampling primitives
\[
\gamma_b=\sum_r|\eta_{b,r}|,\qquad
p_{b,r}=\frac{|\eta_{b,r}|}{\gamma_b},\qquad
s_{b,r}=\mathrm{sign}(\eta_{b,r}).
\]
\end{enumerate}
Set the global overhead $\Gamma=\prod_{b\in\mathcal{B}}\gamma_b$.

\vspace{0.4em}
\textbf{4. Sample mitigated circuits (Monte Carlo).}
Draw $N$ samples. For each sample $k$:
\begin{enumerate}
\item[(a)] For each block $b$ in circuit order, sample an index $r\sim p_{b,r}$ and insert the corresponding physical correction $\widetilde{\mathcal{O}}_{b,r}$ at that block location.
\item[(b)] Record the sample sign $s_k=\prod_{b\in\mathcal{B}} s_{b,r(b)}$.
\item[(c)] Deduplicate by grouping identical sampled circuits $\widetilde C$ and accumulating their multiplicities $m(\widetilde C)$ and net signs $s(\widetilde C)$.
\end{enumerate}

\vspace{0.4em}
\textbf{5. Execute unique sampled circuits.}
For each unique sampled circuit $\widetilde C$, estimate $r(\widetilde C)\approx \langle \hat O \rangle_{\widetilde C}$ using $S$ shots on the QPU (or simulator).

\vspace{0.4em}
\textbf{6. Reweight and combine.}
Return
\[
E=\frac{\Gamma}{N}\sum_{\widetilde C} m(\widetilde C)\,s(\widetilde C)\,r(\widetilde C).
\]

\end{algorithm*}

To address this limitation, we introduce \textsc{MoSAIC} (Modular Spatio-temporal Aggregation for Inverted Channels), a scalable error-mitigation framework that applies PEC at the level of coarse-grained circuit blocks rather than individual noisy layers. As outlined in Algorithm~\ref{algo:mosaic_recipe}, \textsc{MoSAIC} proceeds in  the follwing three stages. 
First, it partitions the transpiled circuit into noise-aligned blocks,
respecting circuit dependencies, hardware locality, and correlated error
processes such as crosstalk. Second, it performs a classical variational optimization to fit a single effective noise channel for each block. This step uses measurement data already available from device characterization and requires no additional quantum tomography. Third, it inverts each learned block channel into a quasi-probabilistic mixture of implementable operations and inserts exactly one recovery per block. Measurement
outcomes are then recombined by classical post-processing to yield an unbiased estimator of the target observable~\cite{PhysRevLett.119.180509,Endo_2018}. This design preserves the unbiasedness of PEC while substantially reducing both sampling overhead and the number of correction insertions across the circuit.

The block size in \textsc{MoSAIC} is a user-tunable knob that trades classical learning cost against quantum sampling cost: larger blocks increase the cost of fitting and channel inversion but reduce the total PEC overhead on hardware. Crucially, as
we show in Sec.~\ref{sec:ta}, the sampling advantage of blockwise over layerwise recovery compounds exponentially with circuit depth, so even modest block sizes yield decisive gains at scale.

Two independent mechanisms drive this exponential advantage:
\begin{enumerate}
    \item \textbf{Inherent noise averaging and cancellation:} Aggregating multiple layers suppresses stochastic fluctuations and partially cancels incoherent error components, yielding an effective noise channel with reduced error strength.
    \item \textbf{Reduced sampling overhead:} Since PEC sampling cost scales exponentially with the number of inverse-channel insertions, applying a single inverse per block, instead of per layer, substantially reduces the overall sampling overhead and extends the practical operating regime of unbiased mitigation. 
\end{enumerate}

We emphasize that \textsc{MoSAIC} is not simply another form of circuit partitioning. Circuit cutting and related partitioning techniques have primarily been developed to distribute large circuits across smaller hardware or to reduce classical simulation cost~\cite{PhysRevLett.125.150504}; they have not been used as a general mechanism for scalable end-to-end quantum error mitigation.
Prior work has shown that recovery operations can sometimes be grouped across an entire circuit in special settings, most notably for Clifford circuits~\cite{CliffordReducedsampling}. However, Clifford circuits are efficiently classically simulable, while practical quantum simulation and variational algorithms typically involve circuits far from Clifford, which is the regime we target with \textsc{MoSAIC}. More recently, Ref.~\cite{Block-PEC} proposed block-based PEC on restricted subcircuits, but under strong, restrictive assumptions such as biased or dephasing-dominated noise and compatibility conditions that allow corrections to commute through a block. In contrast, \textsc{MoSAIC} does not assume any special circuit structure or restricted device noise: it variationally learns an aggregated block noise channel directly and then inverts that learned channel. This makes \textsc{MoSAIC} more general and applicable across a wide range of realistic hardware and algorithmic settings.

In this work, we validate our proposed \textsc{MoSAIC} algorithm on a physically meaningful benchmark: preparation of the critical one-dimensional transverse-field Ising model (TFIM) ground state on IBM's 156-qubit Heron processor, with system sizes up to 50 qubits. Across all tested system sizes, we show that \textsc{MoSAIC} successfully mitigates errors and recovers near-ideal energies at $N=14$ and $N=30$, and at $N=50$. However, under the same sampling budget, standard PEC offers only limited improvements, and its performance deteriorates with increasing system size. 
We further support our hardware results with CUDA-Q simulations on random and variational circuits under a range of noise models. Together, these results show that \textsc{MoSAIC} extends the practical operating regime of unbiased error mitigation on current large-scale superconducting hardware.

\begin{figure*}[t]
    \centering
    \includegraphics[width=\linewidth]{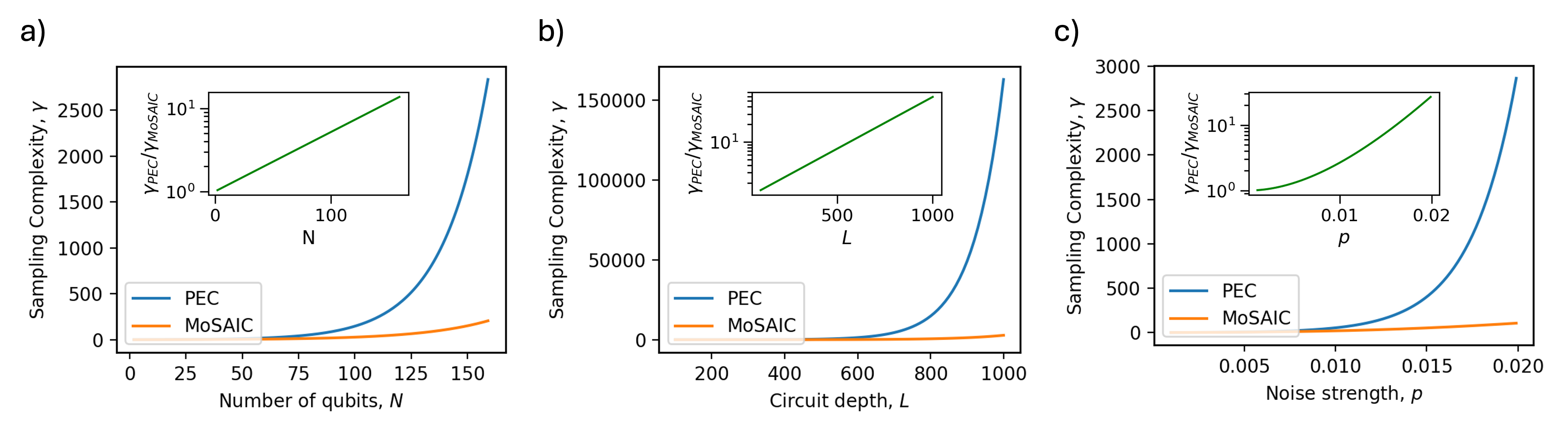}
    \caption{
    \textbf{Sampling-overhead scaling of PEC vs \textsc{MoSAIC}.}
    Sampling overhead factor $\gamma$ for standard PEC and \textsc{MoSAIC} as a function of
    (a) system size $N$, (b) circuit depth $L$, and (c) Pauli noise strength $p$, with the remaining parameters held fixed in each panel.
    Insets show the overhead ratio $\gamma_{\mathrm{PEC}}/\gamma_{\mathrm{MoSAIC}}$ on a logarithmic scale.
    Across all settings, blockwise inversion yields substantially slower growth of $\gamma$ than layerwise PEC, indicating a reduced effective exponent and a rapidly increasing relative advantage with $N$, $L$, and $p$.
    }
\label{fig:overhead_growth}
\end{figure*}

\section{Theoretical Advantages}
\label{sec:ta}

In this section, we formalize why \textsc{MoSAIC} (blockwise PEC) is more scalable than layerwise PEC by showing that the relative advantage of
\textsc{MoSAIC} over standard PEC grows exponentially with system size $N$, circuit depth $L$, and effective noise strength $p$:
\begin{equation}
    \frac{\gamma_{\mathrm{PEC}}}{\gamma_{\mathrm{MoSAIC}}}
    \;\sim\; c_0\,\exp\!\bigl(c_1\,p\,L\,N\bigr),
\end{equation}
where $c_0$ and $c_1$ are constants that depend on the block granularity and noise model. More importantly, applying PEC at the block level is \emph{provably never worse and typically strictly better}
than applying PEC layerwise, in terms of sampling overhead
(proof in Appendix~\ref{appendix:proof}).

To quantify this improvement, we define the \emph{blockwise improvement factor}
\begin{equation}
    \eta \;:=\;
    \frac{\gamma_{\mathrm{layer}}^{\,L}}{\gamma_{\mathrm{block}}},
    \label{eq:eta_def}
\end{equation}
where $L$ is the number of noisy layers per block,
$\gamma_{\mathrm{layer}}$ denotes the PEC sampling overhead when each layer is treated independently,
and $\gamma_{\mathrm{block}}$ denotes the sampling overhead when PEC is applied
\emph{once} to the aggregated block noise channel.
In Appendix~\ref{appendix:proof}, we show that for physically relevant Pauli-type noise models,
$\eta \ge 1$, with strict inequality $\eta > 1$ except in the non-generic case where all contributing
Pauli coefficients share the same sign structure - a condition that does not hold for the physical noise channels considered here.

A concrete calculation for the single-qubit depolarizing channel yields the small-$p$ expansion
\begin{equation}
    \eta
    = \exp\!\left(\frac{L(L-1)}{3}\,p^2 + O(p^3)\right)
    = 1 + \frac{L(L-1)}{3}\,p^2 + O(p^3)
    \label{eqn:eta}
\end{equation}
showing that layerwise PEC incurs a strictly larger sampling overhead,
with the excess cost scaling quadratically in both noise strength and block depth.

Now consider a circuit partitioned into $m_b$ blocks, each of depth $L$.
The total sampling overhead under \textsc{MoSAIC} is
\begin{equation}
    \gamma_{\mathrm{MoSAIC}}
    = \left(\gamma_{\mathrm{block}}\right)^{m_b}.
\end{equation}
In contrast, naive layerwise PEC incurs
\begin{equation}
    \gamma_{\mathrm{PEC}}
    = \left(\gamma_{\mathrm{layer}}^L\right)^{m_b}.
\end{equation}
Therefore, the global improvement factor satisfies
\begin{equation}
    \frac{\gamma_{\mathrm{PEC}}}{\gamma_{\mathrm{MoSAIC}}}
    = \eta^{\,m_b}
    = \left(1 + \frac{L(L-1)}{3}p^2 + O(p^3)\right)^{m_b}.
\end{equation}

Writing $\eta = 1 + \varepsilon$ with $\varepsilon > 0$, demonstrates that even a small per-block advantage compounds across many blocks.
Thus, the improvement factor grows exponentially with the number of blocks (and therefore with
circuit depth), establishing the superior scalability of \textsc{MoSAIC} relative to layerwise PEC.

Figure~\ref{fig:overhead_growth} illustrates these scaling behaviors. Panels (a)--(c) show the dependence of the overhead factor $\gamma$ on (a) the number of qubits $N$, (b) circuit depth $L$,
and (c) noise strength $p$, respectively. In standard PEC, the sampling overhead grows rapidly because a recovery operation is inserted after every
noisy layer. In contrast, MoSAIC inserts recovery operations only at block boundaries, significantly reducing its sampling overhead. The insets show the ratio $\gamma_{\mathrm{PEC}}/\gamma_{\mathrm{MoSAIC}}$
(green curves) on a logarithmic scale, confirming the predicted exponential advantage with $N$, $L$, and $p$.

Taken together, these results show that the benefit of blockwise inversion is not a constant-factor improvement but an exponentially compounding one. Even a modest per-block reduction accumulates across many blocks, making \textsc{MoSAIC} increasingly advantageous as circuits grow in width, depth, and noise level. This reduction in overhead has immediate practical consequences, as it can enable error mitigation at system sizes where standard PEC becomes impractical.

\section{Variational Quantum Eigensolver Benchmark}

In this section, we evaluate \textsc{MoSAIC} on a physically meaningful hardware benchmark: variational ground-state preparation for the one-dimensional transverse-field Ising model (TFIM). 
Variational quantum algorithms are among the most promising near-term applications because they combine quantum state preparation with classical optimization. Among them, the Variational Quantum Eigensolver (VQE)~\cite{TILLY20221, Cerezo2021} is a canonical framework for estimating ground-state energies of many-body Hamiltonians, with applications in quantum chemistry~\cite{McArdle_2020}, materials modeling, and combinatorial optimization. Because VQE circuits accumulate hardware noise with increasing system size and ansatz depth, reliable ground-state preparation provides a
stringent testbed for scalable error mitigation.


A key feature of our protocol is that the VQE optimization is performed
entirely classically via matrix product state (MPS) simulations of the
parameterized circuits. For 1D systems, MPS methods provide essentially exact results up to hundreds of qubits, allowing us to pre-compute high-quality variational parameters for system sizes accessible on current hardware. The pre-optimized circuits can then be deployed directly on our chosen quantum device, creating a compelling near-term workflow. This optimization-then-execute workflow is only viable with robust, scalable error mitigation that preserves the fidelity of the classically optimized state during noisy hardware execution, and we demonstrate that \textsc{MoSAIC} precisely meets this requirement.

\begin{figure*}[t]
    \centering
    \subfloat[Energy-per-site estimates for 14-, 30-, and 50-qubit systems. Standard probabilistic error cancellation (PEC) (orange triangles) provides only partial correction, while \textsc{MoSAIC} (blue diamonds) consistently yields estimates much closer to the ideal values under the same sampling budget. ]{
        \centering
        \includegraphics[width=0.48\textwidth]{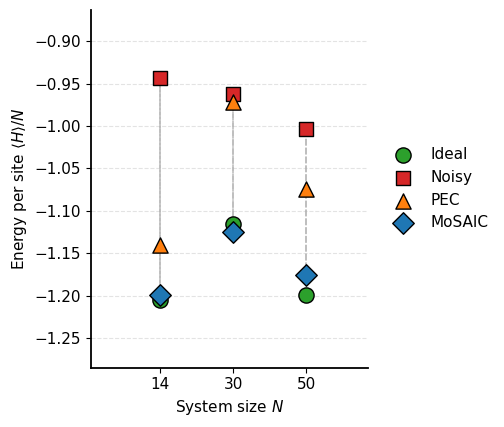}
        \label{fig:num11}
    }
    \hfill
    \subfloat[Variance comparison of mitigation strategies. Shaded error bars denote $\pm 1\sigma$ uncertainty. While \textsc{MoSAIC} maintains relatively small and stable variance across system sizes, PEC exhibits rapidly increasing variance, becoming extremely large at 50 qubits. ]{
        \centering
        \includegraphics[width=0.48\textwidth]{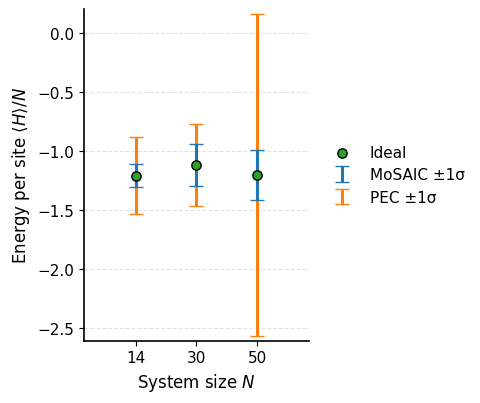}
        \label{fig:num22}
    }
    \caption{VQE benchmark results on the transverse-field Ising model.}
    \label{fig:vqe_mitigation}
\end{figure*}

\subsection{Transverse-Field Ising Model}

This subsection introduces the TFIM benchmark and explains why its critical ground state is a demanding target for both variational preparation and error mitigation.

The one-dimensional transverse-field Ising model is a canonical model of quantum magnetism, with Hamiltonian
\begin{equation}
    H_{\mathrm{TFIM}} = -J \sum_{i} Z_i Z_{i+1} - h \sum_i X_i,
    \label{eq:tfim}
\end{equation}
where \(Z_i\) and \(X_i\) are Pauli operators acting on site \(i\), \(J\) is the Ising coupling strength, and \(h\) is the transverse magnetic field. The model exhibits a quantum phase transition at the critical point \(J=h\), separating a ferromagnetic phase from a paramagnetic phase.

TFIM is a useful benchmark not only because it is theoretically well understood, but also because it captures physically relevant many-body behavior. Many real materials belong to the same universality class as TFIM~\cite{Sachdev1999}, and quasi-one-dimensional compounds such as cobalt niobate are well described by it at low energies~\cite{Coldea2010}. In addition, the GHZ-like states that appear deep in the ferromagnetic phase are relevant for quantum information and metrology applications~\cite{Leibfried2004,Giovannetti2011,RevModPhys.89.035002}.

In this work, we focus on the critical ground state at \(J=h\), where the problem is especially challenging. Near criticality, the spectral gap closes in the thermodynamic limit and the entanglement entropy grows logarithmically with subsystem size, making the state harder to approximate, prepare, and protect from noise. This makes TFIM at criticality an especially strong benchmark for evaluating the scalability of \textsc{MoSAIC}.

\subsection{Variational Ansatz and Optimization}

For the VQE optimization, the cost function is the TFIM energy expectation value
\[
E(\boldsymbol{\theta}) = \langle \psi(\boldsymbol{\theta}) | H_{\mathrm{TFIM}} | \psi(\boldsymbol{\theta}) \rangle,
\]
where \( |\psi(\boldsymbol{\theta})\rangle \) is the parameterized trial state. As mentioned before, we obtain the optimal parameters \(\boldsymbol{\theta}^\star\) fully classically using MPS simulations up to \(N=50\) qubits; further details on the ansatz families and optimized parameters are provided in our GitHub repository~\cite{Yema2025}.

This classical optimization stage is important because it cleanly separates state-design quality from hardware execution quality. The optimized circuit \(U(\boldsymbol{\theta}^\star)\) is executed on hardware without further tuning, so any deviation from the ideal energy can be attributed to device noise and mitigation performance rather than imperfect optimization.

We test \textsc{MoSAIC} on both hardware-efficient and problem-tailored ansatz families to show that its benefit does not depend on a single circuit design. In this way, the benchmark probes the robustness of the mitigation framework across qualitatively different variational circuits.

\subsection{Numerical simulation results}

We first present simulation results that validate \textsc{MoSAIC} before deployment on hardware and show that its mitigation behavior remains stable as system size grows. 

\subsubsection*{Noise Model}
\label{sec:noise}

To emulate realistic device behavior, we consider a correlated Pauli error channel acting on both single- and two-qubit subsystems. Single-qubit Pauli operators \(\{X,Y,Z\}\) are applied with equal probability, and two-qubit Pauli operators \(\{XX,ZZ,XY\}\) are likewise sampled uniformly. We relate the corresponding error strengths by
\begin{equation}
    p_1 = \frac{p_2}{10},
\end{equation}
so that single-qubit errors remain roughly one order of magnitude weaker than two-qubit errors, consistent with typical superconducting-device trends. This model is not intended as an exact reconstruction of a particular processor. Rather, it provides a controlled yet hardware-motivated testbed for assessing how mitigation quality changes with correlated noise strength. For our TFIM numerical simulations, we choose \(p_2=0.06\) to emulate the degree of noise observed on IBM's hardware. 

\subsubsection*{Simulation Results}

\begin{figure*}[t]
    \centering
    \subfloat[Simulated mitigation performance of \textsc{MoSAIC} versus system size for a VQE circuit solving the 1-D transverse-field Ising Hamiltonian. The unmitigated energy per site deviates further from the ideal ground-state value as the number of qubits increases, whereas \textsc{MoSAIC} consistently yields values closer to the ideal baseline, demonstrating scalability under growing circuit width and depth.]{
        \centering
        \includegraphics[width=0.48\textwidth]{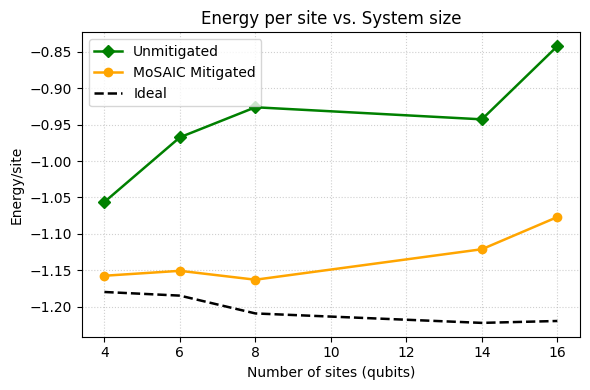}
        \label{fig:num2}
    }
    \hfill
    \subfloat[Simulated mitigation performance of \textsc{MoSAIC} under increasing correlated Pauli noise strength for a 20-qubit random circuit. The observable is $\rho_{00}$, the probability of measuring the all-zero computational basis state. The \emph{Noisy} curve shows degraded accuracy as noise increases, while \textsc{MoSAIC} maintains values close to the ideal reference across all tested noise levels.]{
        \centering
        \includegraphics[width=0.48\textwidth]{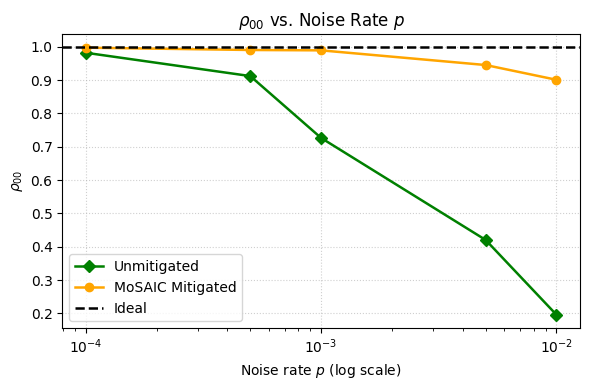}
        \label{fig:num1}
    }
    \caption{Numerical evaluation of \textsc{MoSAIC} on noisy quantum circuit simulations. }
    \label{fig:side-by-side}
\end{figure*}

Figure~\ref{fig:num2} shows the noisy simulation results for system sizes ranging from 4 to 16 qubits - we find that \textsc{MoSAIC} consistently improves the estimated energy across all tested system sizes. The stable mitigation performance as both circuit width and depth increase confirms that blockwise aggregation remains effective beyond small examples and provides a principled basis for real hardware experiments at larger scales.

\subsection{Hardware implementation results}

This subsection compares the noisy hardware execution of the optimized circuits against both standard PEC and \textsc{MoSAIC} mitigated results on the 156-qubit IBM \emph{Heron} processor.

We first transpile our pre-optimized VQE circuits using Qiskit for the native gate set of the Heron r2 processors. We then evaluated the transpiled circuit in three execution modes: (i) unmitigated, (ii) standard PEC using the \texttt{qiskit\_ibm\_runtime} implementation, and (iii) our proposed \textsc{MoSAIC} framework. For reference, we use tensor-network classical simulation as the noise-free \emph{ideal} baseline. No explicit state-preparation-and-measurement (SPAM) correction is applied in any experiment, ensuring that observed improvements are attributable solely to the different mitigation strategies.


\begin{table}[h!]
\centering
\caption{Experimentally measured $\gamma$ for \textsc{MoSAIC} and PEC.}
\vspace{2mm}
\begin{tabular}{lccc}
\hline
\textbf{System size} & \textbf{14 qubits} & \textbf{30 qubits} & \textbf{50 qubits} \\ \hline
$\gamma_{\textsc{MoSAIC}}$ & 1.405 & 3.890 & 6.654 \\
$\gamma_{\textsc{PEC}}$    & 3.549 & 7.771 & 23.044 \\ \hline
\end{tabular}
\label{tab:gamma_comparison}
\end{table}

Figure~\ref{fig:num11} shows the resulting energy-per-site estimates together with their statistical uncertainty in Figure~\ref {fig:num22}. The unmitigated energies deviate from the ideal ground-state values, reflecting the accumulation of both stochastic and coherent hardware errors. For the 14-, 30-, and 50-qubit circuits, the relative errors of the unmitigated estimates are $21.7\%$, $13.8\%$, and $16.3\%$, respectively.
But, \textsc{MoSAIC} can dramatically reduce the relative errors down to $0.46\%$, $0.81\%$, and $1.95\%$ for the 14-, 30-, and 50-qubit cases, respectively, while maintaining tightly controlled statistical uncertainty as shown in Figure~\ref{fig:vqe_mitigation}(b). Notably, these results are obtained despite temporal drift between noise
characterization and circuit execution ranging from several hours to multiple days, confirming the robustness of \textsc{MoSAIC} under realistic operational
conditions.

In contrast, standard PEC provides only limited improvements \emph{under the same sampling budget}. While PEC reduces the bias for smaller systems, its performance deteriorates as circuit size grows: the relative errors for the mitigated energies come out to be $5.35\%$ for 14 qubits, and $12.88\%$ for 30 qubits, and $10.45\%$ for 50 qubits. Moreover, as shown in Figure~\ref{fig:vqe_mitigation}(b), PEC exhibits rapidly increasing variance, with the uncertainty becoming prohibitively large at 50 qubits. 

A complementary perspective is provided by the fraction of device error removed, defined as
\[
1-\frac{|x-\text{Ideal}|}{|\text{Noisy}-\text{Ideal}|}.
\]
Under this metric, \textsc{MoSAIC} removes $97.9\%$, $94.1\%$, and $88.0\%$ of the device error for the 14-, 30-, and 50-qubit systems, respectively. In contrast, PEC removes $75.4\%$, $6.5\%$, and $36.0\%$ of the error.

\begin{table}[h!]
\centering
\caption{Sampling budget used for experiment.}
\vspace{2mm}
\begin{tabular}{lccc}
\hline
\textbf{System size} & \textbf{14 qubits} & \textbf{30 qubits} & \textbf{50 qubits} \\ \hline
\textbf{Samples ($N$)} & 200 & 500 & 1000 \\ \hline
\end{tabular}
\label{tab:sampling_budget}
\end{table}

The effective sampling budgets $N$ used for variance estimation shown in Figure~\ref{fig:vqe_mitigation}(b) are summarized in Table~\ref{tab:sampling_budget}, and are held fixed across mitigation methods for fair comparison.
To quantify the variance difference between \textsc{MoSAIC} and PEC, we relate the sampling variance to the experimentally measured mitigation overhead factor $\gamma$, and show it in Figure~\ref{fig:num22}. The sampling variance of PEC methods scales as 
\[
\sigma^2 \;=\; \frac{\gamma^2}{N},
\]
where $N$ is the effective sampling budget. Table~\ref{tab:gamma_comparison} reports the experimentally extracted $\gamma$ values for both \textsc{MoSAIC} and PEC. For all system sizes, \textsc{MoSAIC} achieves substantially smaller $\gamma$, leading directly to reduced variance. In particular, the overhead gap widens significantly with system size: at 50 qubits, $\gamma_{\textsc{MoSAIC}} = 6.654$ compared to $\gamma_{\textsc{PEC}} = 23.044$, implying an approximate 12x reduction in variance under the same sampling budget. These results highlight a clear scaling advantage: while PEC becomes increasingly ineffective in larger circuits due to variance explosion, \textsc{MoSAIC} maintains both high accuracy and statistical efficiency.

Taken together, these results establish that blockwise probabilistic error cancellation fundamentally alters the scaling of mitigation overhead. By aggregating noise at the block level, \textsc{MoSAIC} strongly weakens the exponential variance growth inherent to layerwise PEC, thereby extending the regime in which unbiased error mitigation remains both accurate and practically usable on current quantum hardware.

\section{Random Circuit Benchmark}

In this section, we evaluate \textsc{MoSAIC} on random circuits to test whether the method remains effective beyond structured physics-inspired ansatz circuits. Although random circuits do not correspond to a specific physical task, they are a demanding stress test for error mitigation because they generate entanglement rapidly and amplify the effect of small early-stage errors on final observables. 

The observable of interest is the probability of measuring the all-zero computational basis state \(\lvert 0 \rangle^{\otimes N}\) at the end of the circuit,
\begin{equation}
\rho_{00} = \langle 0^{\otimes N} \lvert \rho \rvert 0^{\otimes N} \rangle,
\end{equation}
where \(\rho\) is the final simulated density matrix. We use \(\rho_{00}\) as a simple and consistent performance metric across all simulation settings.

Figure~\ref{fig:num1} shows the CUDA-Q accelerated simulation results for a 20-qubit random circuit using the same noise model outlined in Sec~\ref{sec:noise}. It shows that \textsc{MoSAIC} significantly improves the fidelity over the noisy baseline across noise levels spanning roughly two orders of magnitude, and remains close to the ideal value throughout. Note that we do not include standard PEC curves for these larger simulated systems because, in our attempts to do so we found that the layerwise protocol became computationally infeasible beyond $\sim 3$-qubits at the sampling budgets required to achieve substantial mitigation.

To summarize, the TFIM state preparation and random-circuit simulations reinforce the central message of the paper: blockwise aggregation preserves the practical benefits of unbiased mitigation in settings where standard PEC quickly becomes too expensive to use.

\begin{figure*}[t]
    \centering
\includegraphics[width=0.98\textwidth,keepaspectratio]{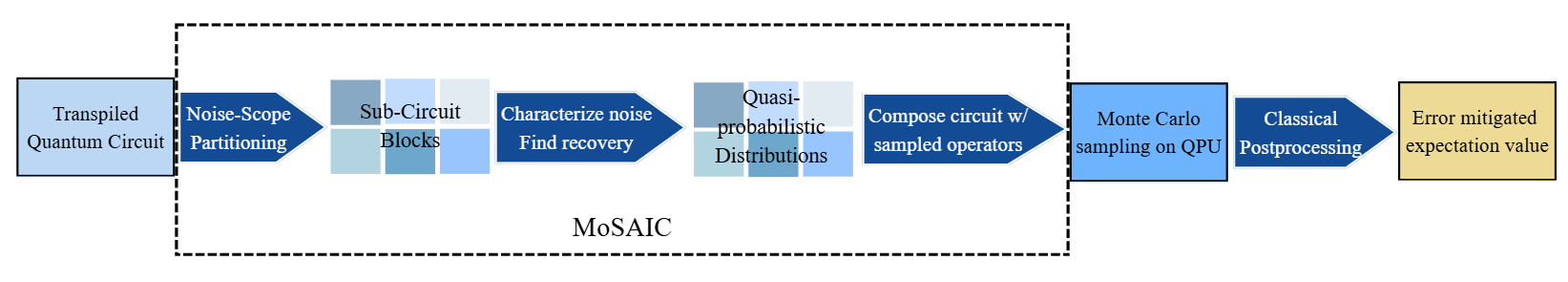}
    \caption{Overview of the \textsc{MoSAIC} error mitigation framework. The workflow proceeds in four main stages: (a) Partitioning: A transpiled quantum circuit is divided into noise-scoped subcircuits (blocks). (b) Characterization: Each block is characterized to learn its effective noise and compute a corresponding quasi-probabilistic recovery distribution. (c) Sampling \& Execution: New composite circuits are formed by sampling from the inverted quasi-probability distributions and are evaluated on the QPU via Monte Carlo sampling. (d) Post-processing: The results are classically combined to yield the final error-mitigated expectation value. Boxed components in the diagram highlight the novel contributions of this work, while the remaining steps adapt the standard PEC procedure with slight modifications.}
    \label{fig:mosaic_overview}
\end{figure*}

\section{Detailed Methodology}
\label{sec:method}

This section describes in detail the full \textsc{MoSAIC} workflow, including noise-aware circuit partitioning, variational block-noise characterization, finding channel inversion, and actual error mitigation procedure.

\textsc{MoSAIC} is a general error-mitigation framework applicable to arbitrary quantum circuits. The overall workflow is illustrated in Figure~\ref{fig:mosaic_overview}, where the boxed components highlight the novel contributions of this work, and the remaining steps follow the standard PEC 
procedure~\cite{van_den_Berg_2023,PhysRevLett.119.180509,PRXQuantum.3.040313,PhysRevA.109.062617} with modifications tailored for \textsc{MoSAIC}. 
Given a target quantum program, we first transpile it to the native gate set of the device.
The resulting circuit is then partitioned into sub-blocks based on noise scope 
and architectural locality. For each circuit block, we variationally estimate an effective noise description using a learned sparse 
Pauli--Lindblad noise model at the layer granularity, and subsequently construct a corresponding 
inverse recovery channel. This inverse channel is decomposed into a quasi-probability distribution over 
implementable Pauli operations, enabling probabilistic sampling of noisy circuit variants. Finally, the 
measurement outcomes across these sampled circuits are recombined through weighted post-processing to 
form an unbiased estimator of the ideal observable. The high-level procedure is summarized in 
Algorithm~\ref{algo:mosaic_recipe}. The remainder of this section explains each of these stages in detail.

\subsection{Noise-Scope Aware Circuit Partitioning}

This subsection explains how \textsc{MoSAIC} partitions a transpiled circuit into smaller blocks that remain faithful to the original computation while exposing useful noise structure.

\begin{figure*}[t]
    \centering
    \subfloat[ Example of noise-scope aware partitioning, where the original circuit is decomposed into subcircuits (blocks) of grain size $w \in \{2,3\}$.]{
        \centering
        \includegraphics[width=0.45\textwidth]{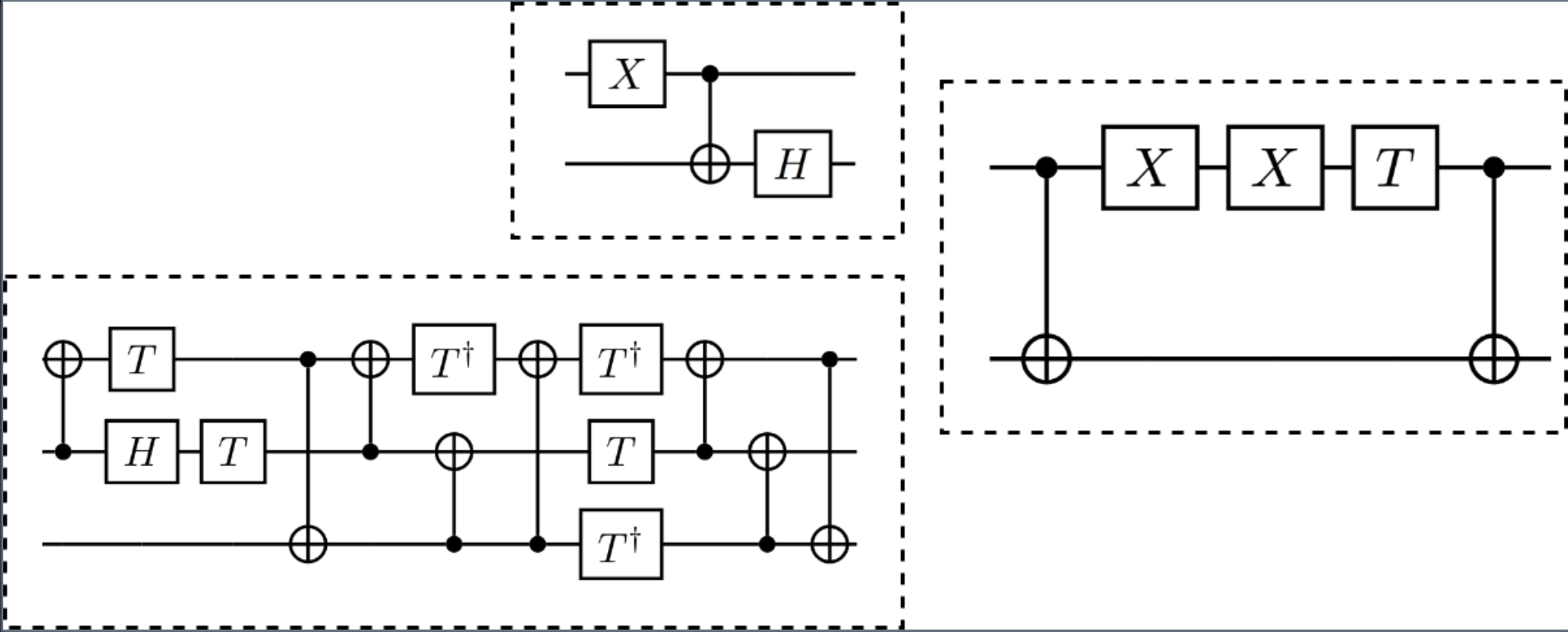}
        \label{fig:partitioning}
    }
    \hfill
    \subfloat[For each block, an effective noise channel 
$B_i$ is variationally characterized, and the corresponding inverse channel $B_i^{-1}$ is appended to 
enable blockwise quasi-probabilistic error cancellation.]{
        \centering
        \includegraphics[width=0.5\textwidth]{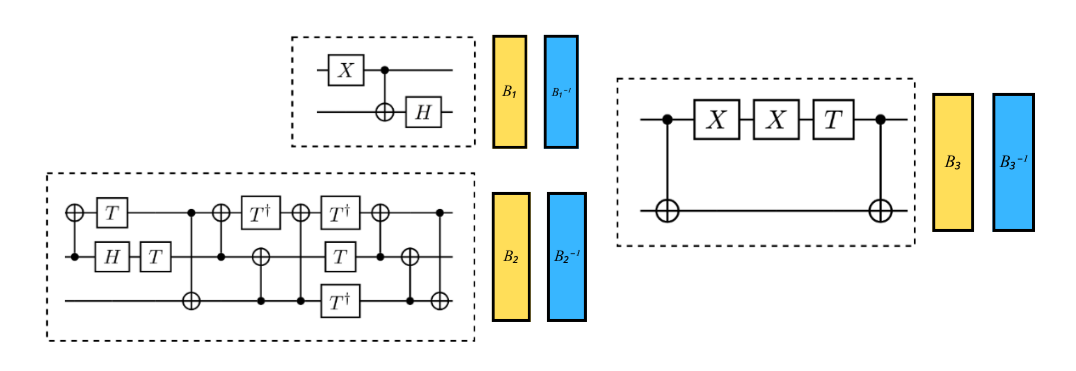}
        \label{fig:recovery}
    }
    \caption{Noise-scope partitioning and corresponding subcircuit characterization.}
\end{figure*}

To enable scalable blockwise PEC, we divide a transpiled circuit into smaller subcircuits whose qubit count and depth are both reduced relative to the full program. Partitioning can be spatial, temporal, or both, allowing the method to capture localized correlated noise while limiting long-range error propagation. An example with block size \(w \in \{2,3\}\) is shown in Figure~\ref{fig:partitioning}. In general, block size need not be uniform across the circuit; heterogeneous block sizing can better match different structural regions of the program.

A key point is that this procedure is not circuit cutting. We do not split gates, insert stitching maps, or change the logical computation. Instead, we use a sequential, non-destructive partitioning strategy that preserves gate order and keeps every operation intact. Blocks are formed heuristically in a single pass over the circuit, with preference given to multi-qubit interactions so that each block captures a meaningful local noise scope while keeping computational volume roughly balanced. The resulting complexity is \(\mathcal{O}(mn)\), where \(m\) is the number of gates and \(n\) is the number of qubits.

The block size \(w\) controls the trade-off between classical preprocessing cost and quantum sampling cost. Larger blocks usually reduce sampling overhead relative to layerwise PEC because more noise is aggregated before inversion, but they also make classical noise learning and quasi-probability decomposition harder, both of which scale exponentially with \(w\). In our Heron VQE experiments, we use \(w=5\) as a practical compromise between classical tractability, mitigation effectiveness, and overall runtime.

In short, the partitioning stage defines the scale at which \textsc{MoSAIC} trades local noise detail for scalable inversion, and it is this balance that makes the later stages computationally viable.

\subsection{Variational Blockwise Noise Characterization}
This subsection describes how \textsc{MoSAIC} propagates layerwise noise channels to a single effective noise channel for each circuit block through optimization.

For each circuit block, we estimate an effective noise channel by aggregating the Pauli--Lindblad noise models~\cite{Breuer2002} extracted at the layer level. For every layer containing two-qubit gates, we obtain a Pauli--Lindblad noise channel $\Lambda$ using the IBM Noise Learner~\cite{van_den_Berg_2023}, parameterized as
\begin{equation}
    \Lambda(\rho) = \exp(\mathcal{L})(\rho) 
    = \prod_{k\in\mathcal{K}}
    \Big( w_k\, \rho + (1 - w_k)\, P_k \rho P_k^\dagger \Big),
    \label{eq:pl_channel_def}
\end{equation}
where $\mathcal{K}$ denotes the set of local Pauli operators $P_k$, $\lambda_k$ are the Lindblad rates, and
$w_k = \tfrac{1}{2}\!\left(1 + e^{-2\lambda_k}\right)$.  
To first order, we approximate $\lambda_k$ as the occurrence probability of error $P_k$, and set the probability of no error to $1 - \sum_{k\in\mathcal{K}} \lambda_k$.

For a block consisting of $N$ layers, the noise superoperator for layer $i$ is denoted
\begin{equation}
    \mathcal{A}_i = \sum_{k} \lambda_k \, (P_k \otimes P_k^{*}),
\end{equation}
and the noisy block superoperator $\mathcal{S}$ is constructed as
\begin{equation}
    \mathcal{S} = 
    \mathcal{A}_N S_N\,
    \mathcal{A}_{N-1} S_{N-1}\cdots
    \mathcal{A}_2 S_2\,
    \mathcal{A}_1 S_1,
\end{equation}
where $S_i$ is the ideal superoperator of the unitary of layer $i$.

Let $U_{\mathrm{block}} = U_N U_{N-1}\cdots U_1$ be the noiseless block unitary.  
We classically learn an effective block noise channel $\mathcal{B}$ such that
\begin{equation}
    \mathcal{B} \otimes \mathcal{B}^{*} \approx 
    \mathcal{S}\left( U_{\mathrm{block}} \otimes U_{\mathrm{block}}^{*}\right)^{-1},
\end{equation}
and solve for $\mathcal{B}$ via the variational optimization problem
\begin{equation}
    \min_{\mathcal{B}} \Big\lVert
    \mathcal{S}\left( U_{\mathrm{block}} \otimes U_{\mathrm{block}}^{*}\right)^{-1} 
    - \mathcal{B} \otimes \mathcal{B}^{*}
    \Big\rVert.
    \label{eq:variational_loss}
\end{equation}

We evaluate several optimizers for solving~\eqref{eq:variational_loss}, including differential evolution, SPSA, and L-BFGS, and find that Adam achieves the best trade-off between runtime and convergence stability. The learning rates and iteration counts are tuned classically for each circuit instance in our VQE experiments and simulations on the IBM Heron processor.

It is important to note that this block-level noise channel learning is performed entirely classically using learned noise model data already collected during execution, requiring no additional on-hardware tomography or calibration, which makes the method both scalable and lightweight in terms of experimental overhead. This variational approach to learning an effective block-level noise channel is also, to our knowledge, not present in prior PEC literature and represents a new pathway for generalized scalable noise learning and inversion.

Once the optimal $\mathcal{B}$ is obtained, we express it as a Pauli channel
\begin{equation}
    \mathcal{B}(\rho)=\sum_{a} c_a\, P_a \rho P_a^\dagger,\qquad
    c_a = \tfrac{1}{2^n}\mathrm{Tr}(P_a^\dagger M),
\end{equation}
where $P_a\in\{I,X,Y,Z\}^{\otimes n}$ for an $n$-qubit block and $M$ is the matrix representation of $\mathcal{B}$.  
The corresponding Kraus operators are $K_a = \sqrt{c_a}\,P_a$, enabling construction of the inverse channel $\mathcal{B}^{-1}$ required for PEC.

We note that the optimization landscape is nonconvex and stochastic, and therefore convergence quality may vary across blocks. In rare cases where the final loss remains large, we repeat optimization with modified hyperparameters or optimizer settings to achieve acceptable accuracy.

\subsection{Channel Inversion}
This subsection explains how the optimized block noise channel is inverted and implemented through quasi-probabilistic sampling.

Probabilistic error cancellation (PEC) relies on applying the inverse of the noise channel to counteract
its effect. In \textsc{MoSAIC}, the inverse channel is obtained from the blockwise noise channel characterized
using the variational procedure described in the previous subsection. Ideally, if a noisy circuit
$\widetilde{\mathcal{C}} = B \circ \mathcal{C}$ is acted on by an inverse recovery channel
$\mathcal{R} = B^{-1}$, then the noise effect can be theoretically removed, yielding an unbiased
estimator of the ideal observable:
\begin{equation}
    B^{-1} \circ \widetilde{\mathcal{C}}
    = \mathcal{R} \circ B \circ \mathcal{C}
    = \mathcal{C}.
    \label{eq:inverse_channel_exact}
\end{equation}

However, although the inverse channel may exist mathematically, it is generally \emph{not physically
implementable}. Noise channels acting in hardware are completely positive and trace preserving
(CPTP), yet by Wigner's theorem~\cite{Wigner1959}, the inverse of an invertible CPTP map is in
general not CPTP. For example, the inverse of an amplitude damping channel is Hermitian preserving
and trace preserving (HPTP), but not completely positive. Consequently, direct implementation of
$B^{-1}$ as a physical quantum operation is impossible in general and instead must be realized
stochastically through quasi-probabilistic sampling.

\vspace{0.5em}
\noindent \textbf{Pauli-channel inversion via Pauli Transfer Matrix (PTM) representation.} For Pauli channels, inversion can be performed efficiently in the PTM representation. In PTM form,
a Pauli channel is diagonal, where each diagonal entry corresponds to an eigenvalue associated with a
Pauli operator. Such a channel is invertible if and only if all eigenvalues are greater than zero, which holds
for any CPTP Pauli channel. Thus, inversion is reduced to inverting a diagonal matrix, followed by
conversion back to a quasi-probabilistic Pauli mixture.

Let $\boldsymbol{\lambda} = (\lambda_0, \lambda_1, \dots, \lambda_{4^n-1})^T$ denote the PTM
eigenvalues for an $n$-qubit block. The corresponding Pauli mixture coefficients
$\boldsymbol{\eta} = (\eta_0, \eta_1, \dots, \eta_{4^n-1})^T$ can be obtained by
\begin{equation}
    \eta_i = \frac{1}{4^n}\sum_{j=0}^{4^n-1} \chi_{ij}\,\lambda_j,
    \qquad i = 0,1,\dots,4^n-1,
\end{equation}
where $\chi_{ij} \in \{+1,-1\}$ is the commutation parity between Pauli operators $P_i$ and $P_j$,
defined by
\begin{equation}
    \chi_{ij} =
    \begin{cases}
        +1, & P_i P_j = +P_j P_i, \\
        -1, & P_i P_j = -P_j P_i.
    \end{cases}
\end{equation}
Equivalently, in matrix form,
\begin{equation}
    \boldsymbol{\eta} = \frac{1}{4^n}\, \chi\, \boldsymbol{\lambda}.
\end{equation}

\vspace{0.5em}
\noindent \textbf{Quasi-probabilistic realization.}
Following the PEC framework, the coefficients $\eta_r$ are converted into a quasi-probability
distribution used for sampling implementable noisy operations. Let $\widetilde{\mathcal{O}}_r$
denote the noisy physical implementation associated with Pauli operator $P_r$. Then,
\begin{equation}
    \mathcal{R}
    = \sum_r \eta_r \widetilde{\mathcal{O}}_r,
    \quad 
    \gamma = \sum_r |\eta_r|,
    \quad
    p_r = \frac{|\eta_r|}{\gamma},
    \quad
    s_r = \frac{\eta_r}{|\eta_r|},
\end{equation}
where $\gamma$ is the sampling overhead factor, $p_r$ is the sampling probability, and $s_r$ is the
classical post-processing sign. Applying $\mathcal{R}$ in this stochastic manner ensures that the
expected outcome of the measurement remains an unbiased estimate of the ideal observable, while the
variance is determined by $\gamma$.

Thus, channel inversion in \textsc{MoSAIC} is obtained exactly in PTM space, and implemented in practice via
quasi-probabilistic sampling, consistent with the PEC framework.

\subsection{Error Mitigation Procedure}
This subsection describes how the blockwise inverse channels are sampled, executed, and recombined into an unbiased mitigated estimator.

Given the quasi-probabilistic representation of the recovery channel
$\mathcal{R} = \sum_r \eta_r\,\widetilde{\mathcal{O}}_r$ obtained from blockwise channel inversion,
the expectation value of an observable $A$ under ideal evolution is
\begin{equation}
\begin{aligned}
\langle A \rangle
&= \operatorname{Tr}\!\left(A\,\rho\right)
= \operatorname{Tr}\!\left(A \sum_{r} \eta_r\, \widetilde{\mathcal{O}}_r(\rho_0)\right) \\
&= \sum_r \eta_r\, \operatorname{Tr}\!\left(A\,\widetilde{\mathcal{O}}_r(\rho_0)\right)
= \gamma\, \sum_r s_r p_r\, \langle A\rangle_r ,
\end{aligned}
\end{equation}
where $\langle A\rangle_r$ denotes the measured observable value from the implementation of
$\widetilde{\mathcal{O}}_r$, $p_r = |\eta_r|/\gamma$ is the sampling probability,
$s_r = \eta_r/|\eta_r|$ is the post-processing sign, and
$\gamma = \sum_r |\eta_r|$ is the sampling overhead factor. Thus, Monte-Carlo sampling over noisy
implementations of $\widetilde{\mathcal{O}}_r$ yields an unbiased estimator of $\langle A\rangle$.

\vspace{0.25em}
\noindent\textbf{Sampling strategy.}
To estimate $\langle A\rangle$, we draw $N$ Monte-Carlo samples, where $N$ scales as
$\mathcal{O}(\gamma^2/\epsilon)$ for desired precision $\epsilon$.
For each sample, we independently draw one Pauli operator per block according to the blockwise
distribution $p_r$. If all sampled operators are identity, the block is left unmodified.
The sub-circuits appended by sampled blockwise operators are then composed into a single executable circuit, and the global
sign is computed as $s = \prod_{b \in \text{blocks}} s_b$, while the global overhead is
$\gamma = \prod_{b \in \text{blocks}} \gamma_b$, shared across all constructed circuits.

\vspace{0.25em}
\noindent\textbf{Circuit deduplication.}
Since many samples may correspond to identical composed circuits (especially when $\gamma$ is small),
we maintain a hash dictionary of unique sampled circuits with occurrence counts $\{m_i\}$.
Each unique circuit is executed once on hardware (or simulation), using $4096$ shots, and its outcome
is recorded as $r_i$. This substantially reduces execution cost while preserving unbiased estimation.

The final mitigated estimator is computed as
\begin{equation}
    E = \frac{\gamma}{N} \sum_{i} s_i\, m_i\, r_i,
\end{equation}
where $N = \sum_i m_i$ is the total number of (non-deduplicated) Monte-Carlo samples.

\vspace{0.25em}
\noindent\textbf{Parallel sampling opportunity.}
For small circuits executed on QPUs with larger available qubit counts, multiple sampled circuits can
be run in parallel on disjoint qubit subsets to further reduce execution time~\cite{aharonov2025reliablehighaccuracyerrormitigation}.  
In our experiments we do not employ parallel sampling, as device characterization (noise learning)
dominated runtime---approximately five times the cost of \textsc{MoSAIC} mitigation per batch. However, when
device characterization is more available or when multiple related circuits are mitigated on the same
hardware, parallel sampling can significantly reduce QPU runtime.

\section{Conclusion and Discussion}

By developing MOSAIC, we demonstrated that the exponential bottleneck traditionally associated with probabilistic error cancellation is not a fundamental limitation. By rethinking the granularity at which noise is modeled and inverted, we can substantially extend the operating regime of unbiased error mitigation. Our experiments on IBM's Heron processor illustrate that \textsc{MoSAIC} can stabilize large-scale, physically relevant VQE circuits well beyond the typical reach standard PEC. In the broader context, \textsc{MoSAIC} points to a promising direction: leveraging classical optimization and architectural noise structure to expand the frontier of experimentally accessible quantum algorithms long before fully fault-tolerant hardware arrives.

Opportunities for future work include (i) improving
the optimization landscape and convergence of blockwise noise learning, potentially using adaptive or
second-order optimizers, Bayesian estimation, or machine-learned noise priors; (ii) exploring dynamic or
heterogeneous block partitioning strategies informed by real-time device diagnostics; (iii) combining
\textsc{MoSAIC} with other mitigation techniques such as virtual distillation, randomized compiling, or
symmetry-based post-selection; and (iv) leveraging hardware parallelism to further reduce sampling
latency. With continued development, \textsc{MoSAIC} represents a promising direction for extending
error-mitigated quantum computation toward utility-scale circuits approaching the 150-qubit regime on
near-term devices.

\begin{acknowledgments}
We thank Zhaohui Yang for helpful discussions. This research used resources of the National Energy Research Scientific Computing Center (NERSC), a Department of Energy Office of Science User Facility under Contract No. DE-AC02-05CH11231 using NERSC award DDR-ERCAP0021493. This research is also
supported by the U.S. National Science Foundation grant CCF-2441912(CAREER), Air Force Office of Scientific Research under award number FA9550-25-1-0146, and the U.S. Department of Energy, Office of  Advanced Scientific Computing Research under Award Number DE-SC0025430. R.J. was also supported by the UCSB NSF Quantum Foundry through Q-AMASE-i program award number DMR-1906325.
\end{acknowledgments}

\bibliographystyle{apsrev4-2}
\bibliography{main}

\appendix

\section{Sampling Overhead of Blockwise PEC}
\label{appendix:proof}

In this appendix, we show that applying probabilistic error cancellation (PEC) to an aggregated
\emph{block noise channel} never incurs larger sampling overhead than applying PEC independently to
each noisy layer. Moreover, under mild and physically relevant conditions, blockwise PEC yields a
\emph{strict} reduction in sampling overhead.

The argument consists of three ingredients:  
(i) submultiplicativity of the PEC sampling overhead under channel composition,  
(ii) invariance of the overhead under ideal unitary conjugation, and  
(iii) destructive interference between quasi-probability coefficients in composed inverse channels.

\subsection{Sampling Overhead and a Normalized Improvement Factor}

Any $n$-qubit Pauli channel admits an expansion
\begin{equation}
    \mathcal{A} = \sum_{i=1}^{4^n} a_i P_i,
\end{equation}
where $\{P_i\}$ is the Pauli operator basis. The PEC sampling overhead associated with $\mathcal{A}$ is
defined as
\begin{equation}
    \gamma(\mathcal{A}) := \sum_{i=1}^{4^n} |a_i|,
\end{equation}
which governs the variance scaling and Monte Carlo sampling cost of unbiased PEC estimators.

To compare blockwise and layerwise PEC, it is convenient to define the \emph{normalized overhead ratio}
\begin{equation}
    \eta(\mathcal{A},\mathcal{B})
    := \frac{\gamma(\mathcal{A}\circ\mathcal{B})}
            {\gamma(\mathcal{A})\,\gamma(\mathcal{B})}.
\end{equation}
By construction, $\eta \le 1$, and $\eta<1$ indicates a strict advantage of applying PEC jointly to
$\mathcal{A}$ and $\mathcal{B}$ rather than separately.

\subsection{Submultiplicativity of Sampling Overhead}

We first establish a general property of $\gamma(\cdot)$.

\paragraph{Lemma (Submultiplicativity).}
For any two Pauli channels $\mathcal{A}$ and $\mathcal{B}$,
\begin{equation}
    \gamma(\mathcal{A}\circ\mathcal{B})
    \le \gamma(\mathcal{A})\,\gamma(\mathcal{B}).
    \label{eq:submult}
\end{equation}

\paragraph{Proof.}
Write
\[
    \mathcal{A}=\sum_i a_i P_i,
    \qquad
    \mathcal{B}=\sum_j b_j P_j.
\]
Since the Pauli group is closed under multiplication,
\[
    \mathcal{A}\circ\mathcal{B}
    = \sum_k c_k P_k,
    \qquad
    c_k = \sum_{i,j} a_i b_j\,\xi_{ijk},
\]
where $\xi_{ijk}\in\{0,1\}$ indicates whether $P_iP_j=P_k$. Applying the triangle inequality yields
\begin{align}
    \gamma(\mathcal{A}\circ\mathcal{B})
    &= \sum_k |c_k|
    \le \sum_{k}\sum_{i,j}|a_i|\,|b_j|\,\xi_{ijk} \\
    &= \sum_{i,j}|a_i|\,|b_j|
    = \gamma(\mathcal{A})\,\gamma(\mathcal{B}).
\end{align}
\hfill$\square$

This inequality guarantees that composing noise channels before applying PEC never increases the
sampling overhead.

\subsection{Invariance under Ideal Unitary Gates}

Ideal unitary channels require no quasi-probabilistic decomposition, and hence
\begin{equation}
    \gamma(\mathcal{U})=\gamma(\mathcal{U}^{-1})=1.
\end{equation}

\paragraph{Lemma (Unitary invariance).}
For any channel $\mathcal{A}$ and unitary $\mathcal{U}$,
\begin{equation}
    \gamma(\mathcal{U}\circ\mathcal{A}\circ\mathcal{U}^{-1})
    = \gamma(\mathcal{A}).
    \label{eq:unitary_invariance}
\end{equation}

\paragraph{Proof.}
By submultiplicativity,
\[
    \gamma(\mathcal{U}\circ\mathcal{A})
    \le \gamma(\mathcal{A}),
\]
and since $\mathcal{A}=\mathcal{U}^{-1}\circ(\mathcal{U}\circ\mathcal{A})$,
\[
    \gamma(\mathcal{A})
    \le \gamma(\mathcal{U}^{-1})\,\gamma(\mathcal{U}\circ\mathcal{A})
    = \gamma(\mathcal{U}\circ\mathcal{A}).
\]
Combining the two inequalities yields equality. \hfill$\square$

As a result, ideal unitary gates may be ignored when analyzing PEC sampling overhead.

\subsection{Blockwise PEC Overhead Bound}

Consider $k$ noisy layers with noise channels $\{\mathcal{A}_1,\dots,\mathcal{A}_k\}$ interleaved with
ideal unitaries. Using unitary invariance~\eqref{eq:unitary_invariance}, the inverse block channel
satisfies
\begin{equation}
    \gamma(\mathcal{A}_{\mathrm{block}}^{-1})
    = \gamma(\mathcal{A}_1^{-1}\circ\cdots\circ\mathcal{A}_k^{-1})
    \le \prod_{i=1}^{k}\gamma(\mathcal{A}_i^{-1}).
\end{equation}
This proves that blockwise PEC never incurs larger sampling overhead than standard layerwise PEC.

\subsection{Generic Strictness from Quasi-Probability Interference}

In practical PEC implementations, inverse channels are represented by quasi-probability expansions
with mixed signs. In particular, for small-noise Pauli--Lindblad models, the inverse channel typically
has the structure
\[
    \eta_I>0,
    \qquad
    \eta_P<0 \quad \text{for all } P\neq I.
\]

When composing two such inverse channels, coefficients associated with non-identity Paulis receive
contributions with opposite signs, leading to partial cancellation. Consequently,
\[
    \left|\sum_{a,b}\eta_a\zeta_b\,\xi_{abk}\right|
    < \sum_{a,b}|\eta_a|\,|\zeta_b|\,\xi_{abk}
    \quad \text{for } P_k\neq I,
\]
and therefore
\begin{equation}
    \gamma(\mathcal{R}_1\circ\mathcal{R}_2)
    < \gamma(\mathcal{R}_1)\,\gamma(\mathcal{R}_2).
\end{equation}
This destructive interference between quasi-probabilities is the fundamental mechanism underlying
the sampling overhead reduction achieved by blockwise PEC.

\subsection{Depolarizing Channel Example}

As an explicit illustration, consider the single-qubit depolarizing channel
\begin{equation}
    \mathcal{E}_p(\rho)
    = (1-p)\rho + p\frac{I}{2}.
\end{equation}
For small $p$, the corresponding PEC overhead satisfies
\begin{equation}
    \gamma(p)
    = 1 + 2p + \frac{8}{3}p^2 + O(p^3).
\end{equation}

For a circuit segment of $L$ identical layers:
\begin{itemize}
    \item \emph{Layerwise PEC} yields
    \[
        \log\gamma_{\mathrm{layerwise}}(L)
        = 2Lp + \frac{2}{3}Lp^2 + O(p^3).
    \]
    \item \emph{Blockwise PEC} yields
    \[
        \log\gamma_{\mathrm{block}}(L)
        = 2Lp + \Bigl(L-\frac{L^2}{3}\Bigr)p^2 + O(p^3).
    \]
\end{itemize}

We define the overhead inflation factor
\begin{equation}
    \eta(L)
    := \frac{\gamma_{\mathrm{layerwise}}(L)}{\gamma_{\mathrm{block}}(L)} .
\end{equation}
Using the above expansions, we obtain
\begin{align}
    \log \eta(L)
    &= \log\gamma_{\mathrm{layerwise}}(L)
       - \log\gamma_{\mathrm{block}}(L) \notag \\
    &= \frac{L(L-1)}{3}p^2 + O(p^3),
\end{align}
and therefore
\begin{equation}
    \eta(L)
    = \exp\!\left( \frac{L(L-1)}{3}p^2 + O(p^3) \right).
\end{equation}
Expanding the exponential for small $p$ yields
\begin{equation}
    \eta(L)
    = 1 + \frac{L(L-1)}{3}p^2 + O(p^3),
\end{equation}
which is strictly greater than unity for any $L>1$.

\medskip
\noindent\textbf{Conclusion.}
Blockwise PEC incurs sampling overhead that is never larger than, and generically strictly smaller
than, the product of layerwise overheads. This provides formal theoretical justification for the
block-level design of \textsc{MoSAIC}. \hfill$\blacksquare$

\end{document}